\documentclass[12pt]{article}

\usepackage{amsthm}
\usepackage{hyperref}
\usepackage{amssymb}
\usepackage{amsmath}
\usepackage{amsfonts}
\usepackage{graphicx}
\usepackage{float}
\usepackage{epstopdf}
\usepackage{caption}
\theoremstyle{plain}

\font\myfont=cmr12 at 10pt

\title{\vspace{-5.0cm}\Large Comparative Study of Two Extensions of Heston Stochastic Volatility Model}
\date{\vspace{-5ex}}
 \author{ \myfont Gifty Malhotra \\ \myfont\href{mailto:giftymalhotradtu@gmail.com}{giftymalhotradtu@gmail.com}
\qquad \and \myfont R. Srivastava \\ \myfont \href{mailto:rsrivastava@dce.ac.in}{rsrivastava@dce.ac.in}
\and \myfont H.C. Taneja \\ \myfont \href{mailto:hctaneja@dce.ac.in}{hctaneja@dce.ac.in}\\
\myfont Department of Applied Mathematics, Delhi Technological University,\\
\myfont Delhi(India)-110042}

\begin{document}

\maketitle
\begin{abstract}
\noindent In the option valuation literature, the shortcomings of one factor stochastic\; volatility\; models\; have\; traditionally\; been\; addressed\; by\; adding jumps to the stock price process. An alternate approach in the context of option pricing and calibration of implied volatility is the addition of a few other factors to the volatility process. This paper contemplates two extensions of the Heston stochastic volatility model. Out of which, one considers the addition of jumps to the stock price process (a stochastic volatility jump diffusion model) and another considers an additional stochastic volatility factor varying at a different time scale (a multiscale stochastic volatility model). An empirical analysis is carried out on the market data of options with different strike prices and maturities, to compare the pricing performance of these models and to capture their implied volatility fit. The unknown parameters of these models are calibrated using the non-linear least square optimization. It has been found that the multiscale stochastic volatility model performs better than the Heston stochastic volatility model and the stochastic volatility jump diffusion model for the data set under consideration.
\smallskip

\end{abstract}
\maketitle

\section{Introduction}
The derivative pricing model proposed by Black and Scholes\cite{black} assumes the volatility to be constant and asset log-return distribution as Gaussian. Empirically, the volatility is not constant but it smiles and the log-return distributions are non-Gaussian in nature characterised by heavy tails and high peaks. A wide range of research has been done to improve upon classical Black-Scholes model. The model has been extended to include either constant volatility with jumps(e.g. jump diffusion (JD) models of Merton\cite{mer76} and Kou\cite{kou02}) or to consider volatility itself as a continuous time stochastic process(e.g. stochastic\; volatility\; models\; given\; by\; Hull and White\cite{hull},\; Scott\cite{scott}, Wiggins\cite{wig}, Stein and Stein\cite{stein}, Heston\cite{heston} and Ball and Roma\cite{ball}, etc.). Stochastic volatility models allow the volatility to fluctuate randomly and are able to incorporate many stylized facts about volatility namely volatility smile and skew, mean reversion and leverage to name a few.\\
\indent In\; the\; single\; factor\; stochastic\; volatility\; models, Heston stochastic volatility model is most popular as it gives a fast and easily implemented semi closed form solution for the European options and is relatively economical from the computational point of view. Despite its success and popularity, it has some shortcomings. The model is unable to fit implied volatility across all strikes and maturities particularity for the options with short expiry \cite{gatheral}.
 Also, Shu and Zhang\cite{shu04} obtained that the Heston model overprices out-of-money (OTM) and short-term options and it underprices in-the-money (ITM) options.\\
\indent In the option valuation literature, the shortcomings of one factor stochastic volatility models have traditionally been addressed by adding jumps to the stock price process(e.g. stochastic volatility jump diffusion (SVJ) models of Bates\cite{bates}, Scott\cite{scot} and Pan\cite{pan}, etc.).
Jumps are added to the stock price dynamics of a stochastic volatility model which improve its pricing performance for the short-term options \cite{bakshi}. An alternate approach is the consideration of multiscale stochastic volatility (MSV) models to address the shortcomings of one factor stochastic volatility models (see \cite{multibook,gifty}). 
In these models, volatility is driven by several factors varying at different time scales. Alizadeh et al.\cite{ali}\; found\; the\; evidence\; of\; two\; factors\; of\; volatility\; with\; one\; highly persistent factor and other quickly mean reverting factor. Extending this idea, Fouque et al.\cite{multi} proposed a two factor\; stochastic\; volatility\; model\; with\; one\; fast\; mean\; reverting\; factor\; and\; another\; slowly\; varying\; factor.
 Christoffersen et al.\cite{chris} empirically showed that the two-factor models improve one factor models in the term structure dimension as well as in the moneyness dimension.\\ 
\indent As both type of models (SVJ or MSV) are the extensions of classical single factor stochastic volatility models, this motivated us to study and compare these two approaches in context of Heston stochastic volatility model.
For this, we have considered two extensions of Heston stochastic volatility model. One is the stochastic volatility jump diffusion model proposed by Yan and Hanson\cite{yan06} which is an extension of Heston stochastic volatility model by adding jumps to the stock price process with log-uniformly distributed jump amplitude.
The another model is the multiscale stochastic volatility model proposed by Fouque and Lorig\cite{fouque11}, in which a fast mean-reverting factor is additionally considered in the framework of Heston stochastic volatility model.
These two models are compared with each other, and also with the Heston stochastic volatility model using S$\&$P $500$ index options data. Firstly, the model parameters are calibrated using non-linear least square optimization.
Then the models' fit to the market implied volatility is captured against log moneyness at different time to maturity. The mean relative error of models' prices with market data is also calculated. We have obtained that the multiscale stochastic volatility model performs better than the other two models.\\
\indent The rest of the paper is organised as follows: The underlying models has been explained in Section \ref{s2}. The empirical analysis has been conducted in Section \ref{s3}, where the calibration of the models' parameters, models' fit to market implied volatility and mean relative error of model prices with market data has been reported and the results obtained are discussed. The conclusion has been given in Section \ref{s4}.
\section{Models Under Consideration}\label{s2}
Firstly, the two models to be considered for the empirical analysis has been explained.
\subsection{Stochastic Volatility Jump Diffusion Model}\label{s2ss1}
Yan and Hanson \cite{yan06} proposed a SVJ model which considers the log-uniform distribution of the jump amplitudes in the stock price process. The model is explained below:\\
\indent Let $X_t$ be the stock price at time $t$ whose dynamics under the risk-neutral probability measure $P^{*}$ is
\begin{equation}\label{e1}
  dX_{t} = X_t((r-\lambda\overline{J})dt + \sqrt {V_t}dW_{t}^{x}+ J(U)dN_{t})
  \end{equation}
where $r$ is the risk free interest rate and  $J(U)$ is the Poisson jump-amplitude with mean $\overline{J}$. The variance $V_t$ follows the CIR process given by
\begin{equation}\label{e2}
  dV_{t} = \kappa(\theta - V_{t})dt + \sigma\sqrt {V_t} dW_{t}^{v}
\end{equation}
with $\kappa$ as the rate of mean-reversion, $\theta$ as the long-run mean value and $\sigma$ as the volatility of variance. The condition $2\kappa \theta \geq \sigma^{2}$ must be satisfied to ensure the positivity of the process \eqref{e2}. $W_{t}^{x}$ and $W_{t}^{v}$ are the standard Brownian motions for the stock price process and the volatility process respectively with correlation
   $$E[dW_{t}^{x}.dW_{t}^{v}] = \rho_{xv} dt$$
$U$ is the amplitude mark process which is assumed to be uniformly distributed with density
\[ \varphi_{U}(u) = \begin{cases} \frac{1}{n-m} & \textnormal{if $m\le u \le n$}, \\ $0$ & \textnormal{otherwise} \end{cases} \]
and is given by
$$ U=\ln (J(U)+1)$$
$N_{t}$ is the standard Poisson jump counting process with jump intensity $\lambda$, $J(U)dN_t$ is the Poisson sum which is given as
$$
J(U)dN_t = \sum_{i=1}^{dN_t}J(U_i)
$$
here $U_i$ is the $i$th jump-amplitude random variable and $\overline{J}$, the mean of jump-amplitude $J$,  is given as
$$
\overline{J}=E[J(U)]= \frac{e^{n}- e^{m}}{n-m}-1.
$$
\indent Under this model, the pricing formula for the European call option, in terms of log stock price $s= \ln(x)$, is given as:
\begin{equation}\label{e3}
C_{svj}=e^{s}P_1(s,v,t,K,T)-Ke^{-r(T-t)}P_2(s,v,t,K,T)
\end{equation}
where $v = V_{t}$ is the variance at time $t$, $T$ is the maturity time, $K$ is the strike price and $r$ is the risk free interest rate. The subscript $svj$ in the price $C_{svj}$ is just to specify the price obtained from SVJ model. The same convention is also followed for Heston and MSV model.\\
For $j=1,2$,
\begin{equation}\label{e4}
P_j(s,v,t,K,T)= \frac{1}{2}+\frac{1}{\pi}\int_{0}^{\infty} Re\biggl[\frac{e^{-i\phi\ln K}f_{j}(s,v,t,\phi,T)}{i\phi}\biggr]d\phi
\end{equation}
where the characteristic function $f_j$ of $P_j$ is
\begin{equation}\label{e5}
f_{j}(s,v,t,\phi,t+\tau)= e^{A_{(1j)}(\tau,\phi)+A_{(2j)}(\tau,\phi)v+i\phi s + \beta_{j}(\tau)}
\end{equation}
with $\tau = T - t$ and $\beta_{j}(\tau)= r\tau\delta_{j,2}; \;\; \delta_{j,2} = 1$ for $j = 2$ and $0$ for $j = 1$.
The other terms are
\begin{align}\label{e6}
&\nonumber \hspace{-0.5cm} A_{(1j)}(\tau,\phi)= r\phi i \tau -(\lambda\overline {J}i \phi+\lambda\overline {J}\delta_{j,1}+r\delta_{j,2})\tau+\lambda\tau\biggl[\frac{e^{(i\phi+\delta_{j,1})n}
-e^{(i\phi+\delta_{j,1})m}}{(n-m)(i\phi+\delta_{j,1})}-1\biggr]\\&\hspace{2cm}+A_{(1j)}^{'}(\tau,\phi)
\end{align}
\begin{equation}\label{e8}
A_{(2j)}(\tau,\phi)=\frac{b_{j}-\rho\sigma\phi i+ d_{j}}{\sigma^{2}}\biggl(\frac{1-e^{d_{j}\tau}}{1-g_{j}e^{d_{j}\tau}}\biggr)
\end{equation}
and
\begin{equation}\label{e7}
A_{(1j)}^{'}(\tau,\phi) = \frac{\kappa \theta}{\sigma^{2}}\biggl[(b_{j}-\rho\sigma\phi i+ d_{j})\tau - 2\ln \biggl(\frac{1-g_{j}e^{d_{j}\tau}}{1-g_{j}}\biggr)\biggr]
\end{equation}
with
$$
g_{j} = \frac{b_{j}-\rho\sigma\phi i+ d_{j}}{b_{j}-\rho\sigma\phi i- d_{j}}
$$
$$
d_{j} = \sqrt{(\rho\sigma\phi i-b_{j})^{2}-\sigma^{2}(2\alpha_{j}\phi i - \phi^{2})}
$$
and
\begin{equation}\label{e9}
\alpha_{1} = \frac{1}{2},\; \alpha_{2} = \frac{-1}{2},\; b_{1}=\kappa-\rho\sigma, \; b_{2}= \kappa
\end{equation}
The unknown parameters of this model are $\kappa, \; \theta, \; \sigma, \; \rho, \; v, \; \lambda, \; m$ and $n$.\\
\indent After the SVJ model, the MSV model of Fouque and Lorig \cite{fouque11} is given below.

\subsection{Multiscale Stochastic Volatility Model}\label{s2ss2}
\noindent Fouque and Lorig \cite{fouque11} extended the Heston stochastic volatility model to a MSV model by considering an additional fast mean-reverting volatility factor in the Heston stochastic volatility model. This model is given below.\\
\indent Under $P^{*}$, the dynamics of stock price $X_t$ is given as
\begin{equation}\label{e10}
  dX_{t} = rX_{t}dt + \eta_{t}X_{t}dW_{t}^{x}
 \end{equation}
here $\eta_{t}= \sqrt{V_{t}}f(Y_t)$. $Y_{t}$ and $V_{t}$ are respectively the fast and the slow scale factors of volatility with their dynamics given as
\begin{equation}\label{e11}
   dY_{t} = \frac{V_t}{\epsilon}(m - Y_{t})dt + \mu\sqrt{2}\sqrt {\frac{V_t}{\epsilon}} dW_{t}^{y}
\end{equation}
and
\begin{equation}\label{e12}
  dV_{t} = \kappa(\theta - V_{t})dt + \sigma\sqrt {V_t} dW_{t}^{v}
\end{equation}
$W_{t}^{x}$ ,$W_{t}^{y}$ and $W_{t}^{v}$ are the standard Brownian motions for the stock price process and for the fast and the slow factors of volatility respectively with
$ E(dW^{x}_{t}.dW^{y}_{t})=\rho_{xy} dt, \;
   E(dW^{x}_{t}.dW^{v}_{t})=\rho_{xv} dt,$ and
    $E(dW^{y}_{t}.dW^{v}_{t})=\rho_{yv} dt $.
$\rho_{xy},\; \rho_{xv}$ and $\rho_{yv}$ are constants which satisfy $\rho_{xy}^{2}<1, \rho_{xv}^{2}<1, \rho_{yv}^{2}<1$ and ${\rho_{xy}^{2}+\rho_{xv}^{2}+\rho_{yv}^{2}-2\rho_{xy}\rho_{xv}\rho_{yv}}<1$.\\
 \indent The fast factor of volatility, $Y_t$ follows the OU process with the mean-reversion rate $V_t/\epsilon$ and volatility of volatility parameter $\mu\sqrt{2}\sqrt {\frac{V_t}{\epsilon}}$. $\epsilon > 0$ is very small so that $Y_t$ is fast mean-reverting towards its long-run mean $m$. The slow volatility factor $V_t$, as already explained for SVJ model, is the square root process. It slowly reverts to its long-run mean $\theta$.\\
\indent Fouque and Lorig\cite{fouque11}\; used\; the\; perturbation\; technique\; to\; obtain\; the\; expression\; for European call option prices. The asymptotic expansion of price in powers of $\sqrt{\epsilon}$ is given as
\begin{equation}\label{e13}
C^{\epsilon}_{msv}(x,y,v,t)= C_{0}+\sqrt{\epsilon}C_{1}+\epsilon C_{2}+...
\end{equation}
\indent They obtained the first order approximation to the price of the European call option as
$$ C^{\epsilon}_{msv}(x,v,t)\approx C_{0}(x,v,t)+\sqrt{\epsilon}C_{1}(x,v,t) $$
\indent This price approximation is clearly independent of the fast factor of volatility and depends only on the slow volatility factor $v$. The approximated price is perturbed around the Heston price $C_0$ at the effective correlation $\rho_{xv}<f>$, where $<f>$ is the average of $f(y)$ with respect to long-run distribution of the volatility factor $Y_{t}$.\\
\indent The first order approximation term $C_1$ is
\begin{equation}\label{e14}
C_{1}=e^{s}Q_1(s,v,t,K,T)-Ke^{-r(T-t)}Q_2(s,v,t,K,T)
\end{equation}
where $s = \ln x$. For $j=1,2$
\begin{equation}\label{e15}
Q_j(s,v,t,K,T)= \frac{1}{2}+\frac{1}{\pi}\int_{0}^{\infty} Re\biggl[\frac{e^{-i\phi\ln K}q_{j}(s,v,t,\phi,T)}{i\phi}\biggr]d\phi
\end{equation}
The characteristic function $q_j$ of $Q_j$ is
\begin{equation}\label{e16}
q_{j}(s,v,t,\phi,t+\tau) = (\kappa \theta \hat{q_{0}}(\tau,\phi)+v \hat{q_{1}}(\tau,\phi))(e^{A_{(1j)}^{'}(\tau,\phi)+A_{(2j)}(\tau,\phi)v+i\phi s})
\end{equation}
here
  \begin{align*}
  & \hat{q_{0}}(\tau,\phi) = \int_{0}^{\tau}\hat{q_{1}}(z,\phi)dz,\\&
 \hat{q_{1}}(\tau,\phi) = \int_{0}^{\tau}B(z,\phi)e^{A_{(3j)}(\tau,\phi,z)}dz
 \end{align*}
 with
 \begin{align}\label{e17}
 &\nonumber A_{(3j)}(\tau,\phi,z) = (b_{j}-\rho\sigma\phi i+ d_{j})\frac{1-g_{j}}{d_{j}g_{j}}\ln \biggl(\frac{g_{j}e^{d_{j}\tau}-1}{g_{j}e^{d_{j}z}-1}\biggr)\\&\nonumber
 B(\tau,\phi) = -(V_{1}A_{(2j)}(\tau,\phi)(2\alpha_{j}\phi i - \phi^{2}) + V_{2}A_{(2j)}^{2}(\tau,\phi)(\phi i)+ V_{3}(2\alpha_{j}\phi^{3}i  + \phi^{2}) \\& \;\;\;\;\;\;\;\;\;\;\;\;\;\;\;\;\;+ V_{4}A_{(2j)}(\tau,\phi)(-\phi^{2}))
 \end{align}
 \indent All the other terms are already given in Eq.\eqref{e7} to Eq.\eqref{e9}. The unknown parameters of this model are $\kappa, \; \theta, \; \sigma, \; \rho, \;v , \; V_{1}, \; V_{2}, \; V_{3}$ and $V_{4}$.\\
 \indent In the next section, the empirical analysis is conducted to compare these models.

\section{\hspace{-0.2cm}Empirical Analysis and Discussion of Results}\label{s3}
 \noindent For the empirical analysis, the data\footnote{Data sharing is not applicable to this article as no new data were created or analyzed in this study.} of S$\&$P $500$ index options is considered from January $4, \; 2016$ with maturity ranging from $30$ days to $180$ days and moneyness from $75\%$ to $125\%$. The risk free rate of interest is the yield on $3$-month U.S. government treasury bill. Firstly, the unknown parameters of each model are calibrated using non-linear least square optimization. Once the parameters are obtained, the models' fit to the market implied volatilities for the S$\&$P $500$ index are captured and plotted against log moneyness. To compare the pricing performance, the mean relative error of each model price is calculated corresponding to the market option price data. These methods are explained in following subsections.
 \subsection{Calibration of Model Parameters}
\begin{figure}[!htbp]
\centering
\hspace*{-2.5cm}
\includegraphics[width=26.5cm]{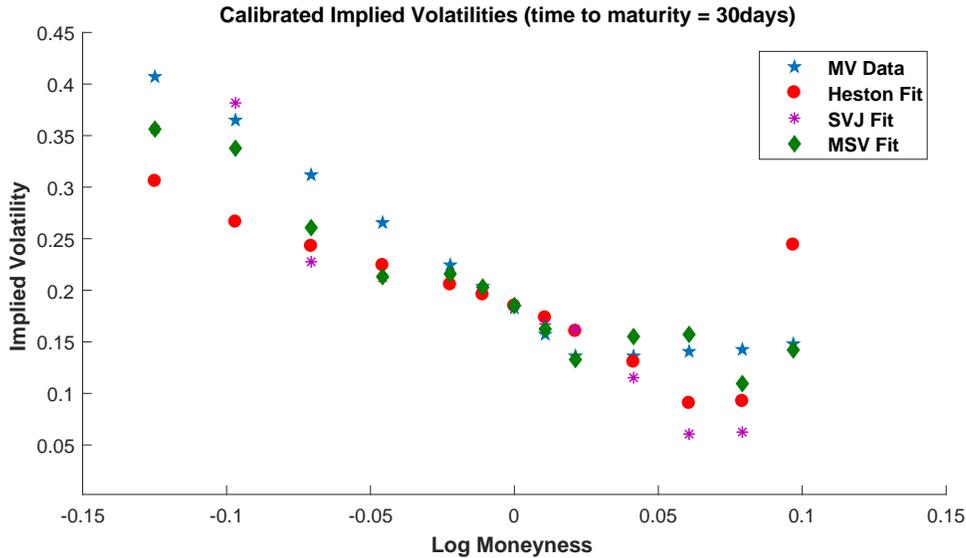}
\vspace*{2.4cm}
\caption{ Models' fit to the implied volatilities of S$\&$P $500$ index with $30$ days to maturity}
\label{f1}
\end{figure}

\begin{figure}[!htbp]
\centering
\hspace*{-2.5cm}
\includegraphics[width=26.5cm]{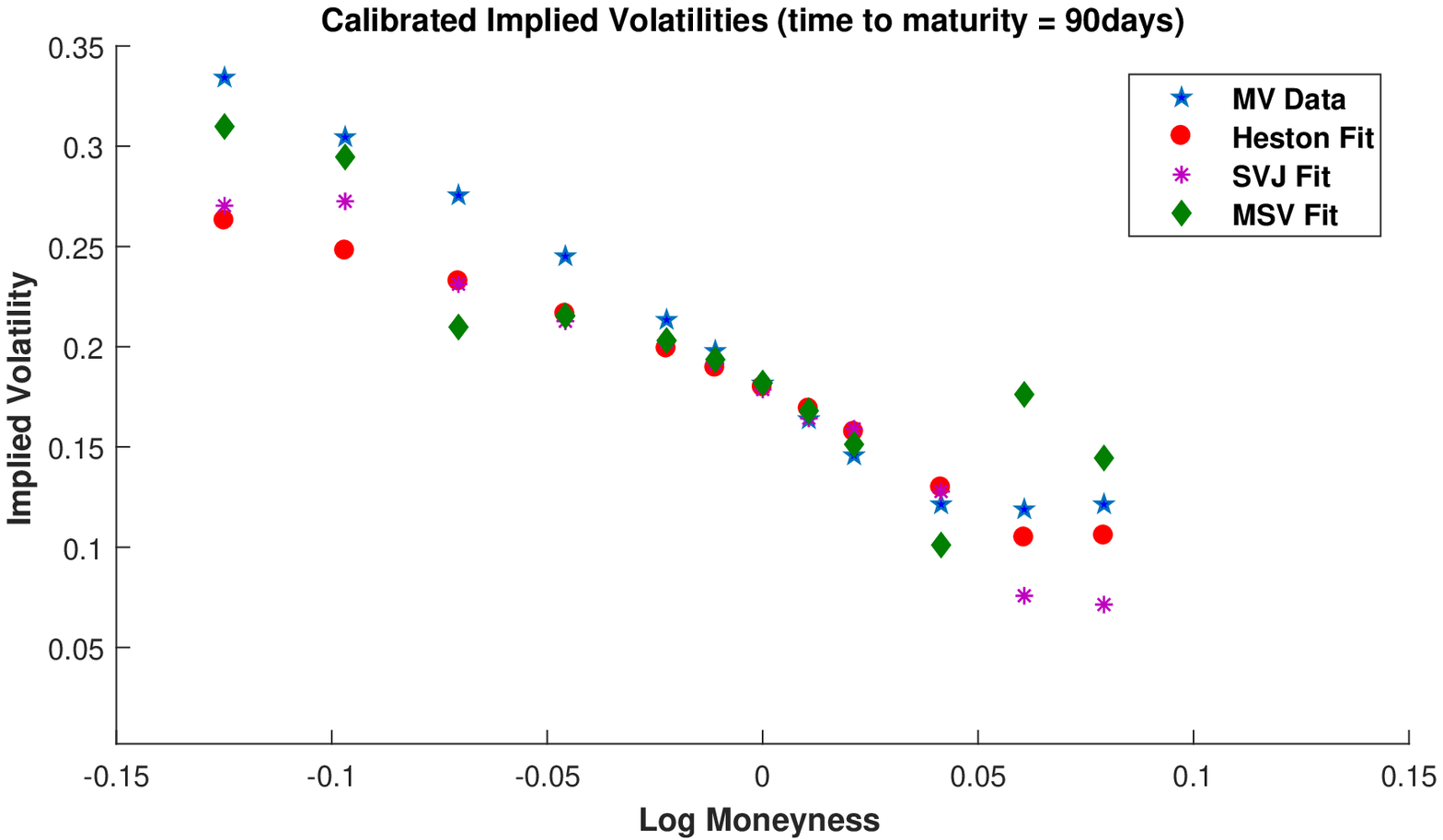}
\vspace*{2.4cm}
\caption{ Models' fit to the implied volatilities of S$\&$P $500$ index with $90$ days to maturity}
\label{f2}
\end{figure}

\begin{figure}[!htbp]
\centering
\hspace*{-2.5cm}
\includegraphics[width=26.5cm]{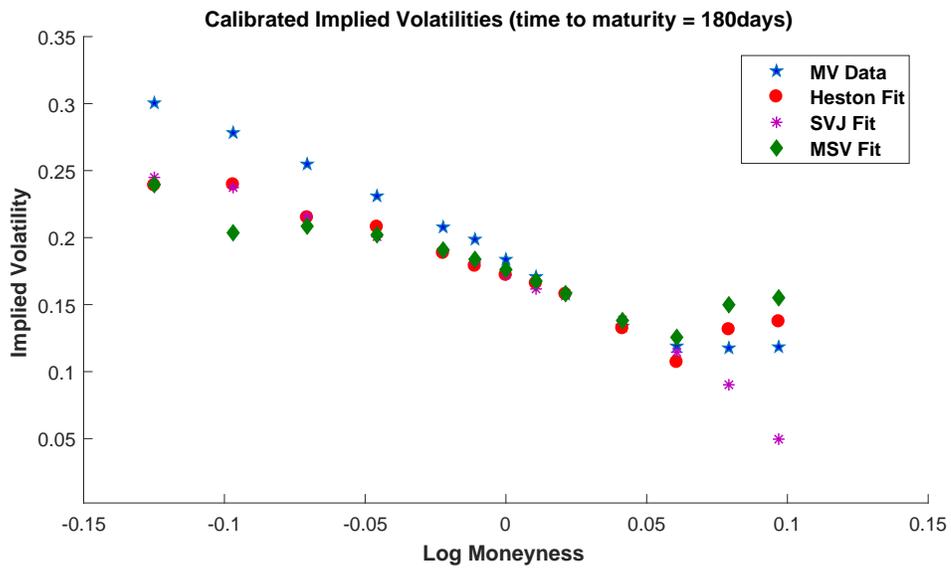}
\vspace*{2.4cm}
\caption{ Models' fit to the implied volatilities of S$\&$P $500$ index with $180$ days to maturity}
\label{f3}
\end{figure}

 \noindent The unknown parameters of Heston stochastic volatility model, SVJ model and MSV model are calibrated using the data of S$\&$P $500$ index options. Let $M_1$, $M_2$ and $M_3$ denote the parameter sets of unknown parameters of Heston, SVJ and MSV model respectively
 , such that
 \begin{align}\label{e18}
 & \nonumber M_1 = (\zeta, \rho, \sigma, \theta, v)\\&\nonumber
 M_2 = (\zeta, \rho, \sigma, \theta, v, \lambda, m, n)\\&
 M_2 = (\zeta, \rho, \sigma, \theta, v, V_1, V_2, V_3, V_4)
 \end{align}
 here all of these unknown parameters are already mentioned in Section \ref{s2} except $\zeta$,
 which is obtained from the condition $2\kappa\theta \geq \sigma^{2}$ of the CIR process \eqref{e2} such that $\zeta = \kappa - \frac{\sigma^{2}}{2\theta},\; \zeta \geq 0$ . Thus, the rate of mean-reversion $\kappa = \zeta + \frac{\sigma^{2}}{2\theta}$ is obtained from the calibrated values of $\zeta,\; \sigma$ and $\theta$. \\
 \indent These parameters are calibrated by non-linear least square optimization using $ \it {MATLAB} 2012b$. The objective function is defined as:
\begin{equation}\label{e19}
\Delta_{h}^{2}(M_1) = \sum_{j}\sum_{i(j)}(C_{mkt}(T_{j},K_{i(j)}) - C_{h}(T_{j},K_{i(j)},M_1))^{2}
\end{equation}
\begin{equation}\label{e20}
\Delta_{svj}^{2}(M_2) = \sum_{j}\sum_{i(j)}(C_{mkt}(T_{j},K_{i(j)}) - C_{svj}(T_{j},K_{i(j)},M_2))^{2}
\end{equation}
\begin{equation}\label{e21}
\Delta_{msv}^{2}(M_3) = \sum_{j}\sum_{i(j)}(C_{mkt}(T_{j},K_{i(j)}) - C_{msv}(T_{j},K_{i(j)},M_3))^{2}
\end{equation}
 where $C_{mkt}(T_{j},K_{i(j)})$ is the market price of call option with maturity $T_{j}$. For each expiration $T_{j}$, the available collection of strike prices is $K_{i(j)}$. Similarly, for a particular value of $T_{j}$ and $K_{i(j)}$, $C_{h}(T_{j},K_{i(j)},M_1)$, $C_{svj}(T_{j},K_{i(j)},M_2)$ and $C_{msv}(T_{j},K_{i(j)},M_3)$ are the prices of the European call options with expiration date $T_{j}$\; and\; exercise\; price\; $K_{i(j)}$,\; calculated\; from\; the\; Heston\; stochastic\; volatility model with parameter set $M_1$, SVJ model with the parameter set $M_2$ and MSV model with the parameter set $M_3$ respectively.\\
 \indent The optimal set of parameters $M_{1}^{*}$,$M_{2}^{*}$ and $M_{3}^{*}$ is obtained which satisfies
 \begin{align}
 & \nonumber \Delta_{h}^{2}(M_{1}^{*}) = min_{M_1} (\Delta_{h}^{2}(M_1))\\&\nonumber
\Delta_{svj}^{2}(M_{2}^{*})= min_{M_2} (\Delta_{svj}^{2}(M_2))\\&
\Delta_{msv}^{2}(M_{3}^{*})= min_{M_3} (\Delta_{msv}^{2}(M_3))
\end{align}
\indent Firstly, the optimal parameter set for the Heston stochastic volatility model, $M_{1}^{*}$, is calibrated.
Once the $M_{1}^{*}$ is obtained, the initial iteration for SVJ model is taken as $(M_{1}^{*}, \;50, \;-0.01, \;0.01)$ with the lower and upper bounds\; for\; the\; last\; three\; components\; as\; $(1, \;-1, \;0)$\; and\; $(100, \;0, \;1)$\; respectively. Similarly, the\; initial\; iteration\; for\; MSV\; model\; is\; taken\; as\; $(M_{1}^{*}, \;0.0001, \;0, \;0, \;0)$ with \; the \; lower\; and\; upper\; bounds\; for\; last\; four\; components\; as $(-0.05, \;-0.05, \;-0.05, \;-0.05)$ and $(0.05, \;0.05, \;0.05, \;0.05)$ respectively.\\
\indent Using the optimal parameter set, the implied volatility fit for all the three models is obtained and is plotted against log moneyness ($log \frac{K}{X}$). The models fit are compared relative to market implied volatility (MV) data. It is given in Fig.\ref{f1} to Fig.\ref{f3} for time to maturity $30$ days, $90$ days and $180$ days respectively.\\
\indent The parameters are calibrated from the whole data but the results are given and discussed for the different maturity times, separately.\\


\indent Along with this, the mean relative error of the prices obtained from Heston stochastic volatility model, SVJ model and MSV model, with the market data of S$\&$P $500$ index option, is calculated at different maturity time. It is given in subsection \ref{s2ss3}.
\subsection{Mean Relative Error}\label{s2ss3}
\noindent For a particular model with price $C_{model}(T_{j},K_{i(j)},\Theta)$ at different values of $T_{j}$ and $K_{i(j)}$, the mean relative error (MRE) of model price with respect to market price, at time to maturity $T_{j}$ is given as
\begin{equation}
MRE(j) = \frac{1}{N_{j}}\sum_{i(j)}\frac{|C_{model}(T_{j},K_{i(j)},\Theta) - C_{mkt}(T_{j},K_{i(j)})|}{C_{mkt}(T_{j},K_{i(j)})}
\end{equation}
where $N_{j}$ is the different number of call options that has expiry at time $T_{j}$, $\Theta$ is the optimal parameter set for the given model.\\
\indent The mean relative error of Heston stochastic volatility model, SVJ model and MSV model is calculated for S$\&$P $500$ index data set. The maturity time is taken from $30$ days to $180$ days. Corresponding to a particular maturity, the strike prices range from $75\%$ to $125\%$. The results are given in Table \ref{t1}.\\
\begin{table}[!htbp]
\vspace{-0.3cm}
\caption{\label{t1} The mean relative error of models prices with respect to market data.}
\centering
\scalebox{1.2}{%
\begin{tabular}{|c|ccc|}
 \hline
 & & Models & \\
  Maturity Time (T)& Heston & SVJ & MSV \\ \hline
   $30$ days & 0.0697 & 0.0499 & 0.0225 \\
  $90$ days & 0.0874 & 0.0987 & 0.0456 \\
  $180$ days & 0.0284 & 0.1070 & 0.0380 \\ \hline
  \end{tabular}}
\end{table}

\indent Now, we discuss the results obtained in
 Fig.\ref{f1} to Fig.\ref{f3} and in Table\ref{t1}. From the models fit to the implied volatility given in  Fig.\ref{f1} to Fig.\ref{f3}, it is clearly observable that the MSV model performs in an improved way in comparison to Heston stochastic volatility model and the SVJ model. For at-the-money (ATM) and near the money options, all the three models give equivalent results. The difference is observable for ITM and OTM options.\\
\indent In Fig.\ref{f1}, the maturity time is short, that is $30$ days. For such options, the Heston model fit to market implied volatility is not good. This supports the empirical findings that the Heston model poorly performs for short term options. The SVJ model performs better than Heston model and MSV model for deep ITM options, but as the log moneyness value increases, MSV model outperforms both Heston model and SVJ model.\\
\indent In Fig.\ref{f2}, the maturity time is medium, that is $90$ days. The implied volatility fit of Heston model is improved for the OTM options. For ITM options, implied\; volatility\; fit\; of\; MSV\; model\; is\; better\; than\; the\; implied\; volatility\; fit\; of\; Heston model. The Heston model fit is equivalent to the SVJ model fit to market implied volatility.\\
\indent In Fig.\ref{f3}, at the longer maturity, which is $180$ days, all of the three models give almost similar fit for ITM options but for OTM options the Heston model outperforms the other two models. The implied volatility fit of MSV model is better than the fit of SVJ model to the market implied volatilities. Thus, out of SVJ model and MSV model, the overall fit of MSV model to the market implied volatility is better than SVJ model.\\
\indent Additionally, from Table\ref{t1}, the pricing performance of three models is compared in terms of mean relative error of models prices with the market option price data. For the short and medium term options with maturity $30$ and $90$ days respectively, the mean relative error of MSV model is least. Thus the MSV model performs better than the SVJ model and Heston model in pricing. For maturity time $30$ days, SVJ model performs better than Heston model in pricing, but for maturity $90$ days, Heston model gives better pricing performance.\\
\indent For the long term options with maturity $180$ days, the MSV model performs better than SVJ model and Heston model outperforms the SVJ and MSV model.\\
\indent Thus, out of SVJ model and MSV model, the overall pricing performance of MSV model is better than SVJ model for the data set under consideration.
\section{Conclusion}\label{s4}
\noindent The two extensions of Heston stochastic volatility model, already proposed in literature, are studied and compared in this paper on the basis of their fit to the market implied volatility and pricing performance. An empirical analysis is conducted on S$\&$P $500$ index options data and the results are obtained for all the three models. It has been obtained that for the data set under consideration,
multiscale stochastic volatility performs better than the stochastic volatility jump model. Thus, the inclusion of additional volatility factor to a stochastic volatility model enhances its fit to the market implied volatility and improves its pricing performance in comparison to the addition of jump factors to the underlying stock price process.

\end{document}